\documentclass[aps,prc,letterpaper,11pt,twoside,tightenlines,nofootinbib,showpacs,preprint,superscriptaddress]{revtex4-1}
\usepackage[colorlinks=true]{hyperref}
\hypersetup{colorlinks=true, citecolor=blue, urlcolor=blue, linkcolor=blue}
\usepackage{epsfig}
\usepackage{subfigure}
\usepackage{amsmath}
\usepackage{rotating}   
\usepackage{xcolor}      
\begin{document}

\arraycolsep1.5pt
\newcommand{\Ima}{\textrm{Im}}
\newcommand{\Rea}{\textrm{Re}}
\newcommand{\mev}{\textrm{ MeV}}
\newcommand{\gev}{\textrm{ GeV}}

\title{Molecular nature of $P_{cs} (4459)$ and its heavy quark spin partners}

\author{C. W. Xiao}
\email{xiaochw@csu.edu.cn}
\affiliation{School of Physics and Electronics, Central South University, Changsha 410083, China}

\author{J. J. Wu}
\email{wujiajun@ucas.ac.cn}
\affiliation{School of Physical Sciences, University of Chinese Academy of Sciences (UCAS), Beijing 100049, China}

\author{B. S. Zou}
\email{zoubs@itp.ac.cn}
\affiliation{Key Laboratory of Theoretical Physics, Institute of Theoretical Physics, Chinese Academy of Sciences, Beijing 100190, China}
\affiliation{School of Physical Sciences, University of Chinese Academy of Sciences (UCAS), Beijing 100049, China}
\affiliation{School of Physics and Electronics, Central South University, Changsha 410083, China}

\date{\today}

\begin{abstract}

Inspired by the observation of the $P_{cs} (4459)$ state by LHCb recently, we reexamine the results of the interaction of the $J/\psi \Lambda$ channel with its coupled channels, exploiting the coupled channel unitary approach combined with heavy quark spin and local hidden gauge symmetries. 
By tuning the only free parameter, we find a pole of $(4459.07+i6.89)\mev$ below the $\bar D^* \Xi_c$ threshold, which was consistent well with the mass and width of the $P_{cs} (4459)$ state. 
Thus, we assume the $P_{cs} (4459)$ state to be a $\bar D^* \Xi_c$ bound state with the uncertainties on its degeneracy with $J^P = \frac{1}{2}^-$ and $J^P = \frac{3}{2}^-$. 
For the degeneracy, it would have two-poles structure, like $P_c (4450)$ before. 
There is another pole in the $J^P = \frac{1}{2}^-$ sector, $(4310.53+i8.23)\mev$, corresponding to a deep bound state of $\bar D \Xi_c$. 
Furthermore, the previously predicted loose bound states of $\bar D \Xi'_c$, $\bar D^* \Xi'_c$, $\bar D^* \Xi^*_c$ with $J=1/2,~I=0$ and $\bar D^* \Xi'_c$, $\bar D \Xi^*_c$, $\bar D^* \Xi_c^*$ with $J=3/2,~I=0$ may exist as either bound states or unbound virtual states.
We hope that future experiments can search for the $\bar D^{(*)} \Xi_c$ molecular states in their dominant decay channels of $\bar D^{(*)}_s  \Lambda_c$,  also in the $J/\psi \Lambda$ and $\eta_c \Lambda$ channels to reveal their different nature. 

\end{abstract}

\maketitle

\section{Introduction}

Early in 2010, with the coupled channel unitary approach (CCUA)~\cite{Oller:1997ti,Oset:1997it,Oller:2000fj} and the local hidden gauge (LHG) formalism~\cite{Bando:1984ej,Bando:1987br,Meissner:1987ge,Nagahiro:2008cv} combined with SU(4) symmetry, several hidden charm and hidden charm strangeness resonances were predicted around the energy range of 4200 MeV -- 4600 MeV in Ref.~\cite{Wu:2010jy}, 
where the possible decay channel of $\eta_c N$ or $J/\psi N$ was suggested for searching for the hidden charm ones in the experiments and its further investigation was given in detail in Ref.~\cite{Wu:2010vk}.
After that, the pentaquark states caught the theoretical interest once again and had many predictions in the hidden charm sector in Refs.~\cite{Wang:2011rga,Yang:2011wz,Yuan:2012wz,Wu:2012md,Garcia-Recio:2013gaa,Xiao:2013yca,Uchino:2015uha,Karliner:2015ina}. 
In 2015, the LHCb Collaboration reported two pentaquark-like resonances found in the $J/\psi p$ invariant mass distributions of the $\Lambda_b^0 \to J/\psi K^- p$ decay~\cite{Aaij:2015tga,Aaij:2015fea}, denoted as $P_c(4380)^+$ with a large width about $205\mev$ and $P_c(4450)^+$ with a small width of $39\mev$, which were confirmed by a model-independent re-analysis of the experimental data \cite{Aaij:2016phn} and in the $\Lambda_b^0 \to J/\psi p \pi^-$ decay \cite{Aaij:2016ymb}. 
Furthermore, in 2019, with the Run-2 data the LHCb Collaboration updated the new results for the $P_c$ states as three clear narrow structures \cite{Aaij:2019vzc},
\begin{align*}
&M_{P_{c1}} = (4311.9\pm0.7^{+6.8}_{-0.6})\mev, \quad
  \Gamma_{P_{c1}}= (9.8\pm2.7^{+3.7}_{-4.5})\mev,  \\
&M_{P_{c2}} = (4440.3\pm1.3^{+4.1}_{-4.7})\mev, \quad
  \Gamma_{P_{c2}}= (20.6\pm4.9^{+8.7}_{-10.1})\mev, \\
&M_{P_{c3}} = (4457.3\pm0.6^{+4.1}_{-1.7})\mev, \quad
  \Gamma_{P_{c3}}= (6.4\pm2.0^{+5.7}_{-1.9})\mev,
\end{align*}
where the $P_c(4450)$ was split into two states of $P_c(4440)$ and $P_c(4457)$, in addition to a new narrow resonance $P_c(4312)$. 
Whereas, the broad old one $P_c(4380)$ could neither be confirmed nor refuted in the updated results.

In principle, due to the fact that the $P_c$ states were found in the experiments, the $P_{cs}$ states as their strangeness partners should also exist as predicted in Refs.~\cite{Wu:2010jy,Wu:2010vk} from SU(4) flavour symmetry. 
Note that, the first predictions for the masses and the widths of the molecular resonances in Refs.~\cite{Wu:2010jy,Wu:2010vk} were due to the lack of the experimental informations to determine the only free parameter of $a_\mu$ ($\mu$ is not an independent one~\cite{Oller:2000fj,Ozpineci:2013zas,Xiao:2020frg}) in the loop functions. 
Therefore, after the updated results of LHCb Collaboration~\cite{Aaij:2019vzc} became available, this free parameter was fitted as $a_\mu (\mu=1\gev) = -2.09$ in Ref.~\cite{Xiao:2019aya} with the masses of three $P_c$ states based on the former work of~\cite{Xiao:2013yca}, where seven hidden charm molecular states were predicted with the CCUA combined with the LHG symmetry and the heavy quark spin symmetry (HQSS)~\cite{Isgur:1989vq,Neubert:1993mb,MW00}. 
Inspired by the experimental findings of the $P_c$ states, with the same fitted parameter of $a_\mu$ from the $P_c$ states, and also combined with the LHG symmetry and the HQSS, several bound states of $\bar{D}^{(*)} \Xi_c^{(*)}$ and $\bar{D}^{(*)} \Xi_c^\prime$ were predicted in Ref.~\cite{Xiao:2019gjd}, where some of them were analogous to the ones in Refs.~\cite{Wu:2010jy,Wu:2010vk} and seeking for them in the reaction of $\Xi_b^- \to J/\psi \Lambda K^-$ was commented at the end as also suggested in the work of~\cite{Chen:2015sxa,Santopinto:2016pkp}. 
The possibility to look for them in the $\Lambda_b$ decays was also discussed in the early work of~\cite{Feijoo:2015kts,Lu:2016roh}.
Furthermore, the decay properties of the hidden charm strangeness resonances predicted in Refs.~\cite{Wu:2010jy,Wu:2010vk} were investigated in detail in Ref.~\cite{Shen:2019evi}, where the partial decay widths of possible decay channels were obtained by exploiting the effective Lagrangian framework through the triangle loops. 
On the other hand, with the one-boson-exchange model, the work of Ref.~\cite{Chen:2016ryt} also predicted that the possible $\bar{D}_s^* \Sigma_c^{(*)}$ and $\bar{D}^* \Xi_c^{(\prime,\,*)}$ pentaquark states existed. 
Taking into account the spin-flavour symmetric states, the possible $SU_f(3)$ multiplets for the charmonium compact pentaquark states were obtained in Ref.~\cite{Santopinto:2016pkp}, where the possible decay channels and the partial decay widths were discussed for these predictions. 
Using the chiral effective field theory up to the next-to-leading order, Ref.~\cite{Wang:2019nvm} found ten bound states in the hidden charm strangeness systems of $\bar{D}^{(*)} \Xi_c^\prime$ and $\bar{D}^{(*)} \Xi_c^{(*)}$ with different spin, where a mass difference of about 6 MeV for two $\bar{D}^* \Xi_c$ molecular states with spins $J=\frac{1}{2}$ and $J=\frac{3}{2}$ was obtained and looking for these $P_{cs}$ states in the decay channel of $J/\psi \Lambda$ was suggested. 

Just recently, the LHCb Collaboration had reported the results of $\Xi_b^- \to J/\psi \Lambda K^-$ decay in Ref.~\cite{Aaij:2020gdg}, where a resonance structure of $P_{cs} (4459)$ state was found in the invariant mass distributions of $J/\psi \Lambda$, given as
\begin{equation*}
M_{P_{cs}} = (4458.8\pm2.9^{+4.7}_{-1.1})\mev, \quad
  \Gamma_{P_{cs}}= (17.3\pm6.5^{+8.0}_{-5.7})\mev,
\end{equation*}
which is just about 19 MeV below the $\bar{D}^{*0} \Xi_c^0$ threshold. 
Note that, analogous to the degenerate $P_c(4450)$ state, the possible hypothetical structure as predicted in Refs.~\cite{Xiao:2019gjd,Wang:2019nvm} with different spins was unsure for the current data sample. 
The two-poles structure was also suggested in the results of the QCD sum rules in Ref.~\cite{Chen:2020uif}, where the $P_{cs}(4459)$ state was assumed to be a $\bar{D}^* \Xi_c$ molecular state with spin-parity $J^P=\frac{1}{2}^-$ or $J^P=\frac{3}{2}^-$ and a $\bar{D} \Xi_c$ bound state at $4.29^{+0.13}_{-0.12} \gev$ with $J^P=\frac{1}{2}^-$ was found too. 
Also with the QCD sum rules, Ref.~\cite{Wang:2020eep} assigned the $P_{cs}(4459)$ state as a hidden-charm compact pentaquark state with $J^P=\frac{1}{2}^-$, which was confirmed in the later work of~\cite{Azizi:2021utt,1844557}. 
On the other hand, with a combined effective field theory and phenomenological assumptions, the work of~\cite{Peng:2020hql} proposed that the $P_{cs}(4459)$ state could be a $\bar{D}^* \Xi_c$ molecular pentaquark state with $J^P=\frac{3}{2}^-$ more likely, or possibly $J^P=\frac{1}{2}^-$ with more uncertainties on its mass. 
Using a coupled channel analysis based on the one-boson-exchange model, Ref.~\cite{Chen:2020kco} concluded that the $P_{cs}(4459)$ state was not a pure $\bar{D}^* \Xi_c$ molecular state and other possible resonances of $\bar{D}^{(*)} \Xi_c^{(\prime,\, *)}$ could exist, and its decay behavior was further discussed in Ref. \cite{Chen:2021tip}. 
In Ref.~\cite{Liu:2020hcv}, the existence of the $\bar{D}^{(*)} \Xi_c^{(\prime,\, *)}$ molecular states were also found under the effective field theory by taking into account the HQSS and SU(3) flavor symmetry, which were the SU(3)-flavour partners of the $\bar{D}^{(*)} \Sigma_c^{(*)}$ molecular states. 
With the quasipotential Bethe-Salpeter equation approach, Ref.~\cite{Zhu:2021lhd} assigned the $P_{cs}(4459)$ state as the $\bar{D}^* \Xi_c$ molecular state with $J^P=\frac{3}{2}^-$ and predicted other $\bar{D}^{(*)} \Xi_c^{(\prime,\, *)}$ states with different spins.
Furthermore, the decay of $\Lambda_b \to J/\psi \Lambda \phi$ was suggested to search for the $P_{cs}(4459)$ state in Ref.~\cite{Liu:2020ajv}.

Motivated by the new findings of Ref.~\cite{Aaij:2020gdg}, we reexamine the results of Refs.~\cite{Wu:2010jy,Wu:2010vk,Xiao:2019gjd} to understand the differences between our predictions and experimental data, then fix the model parameters to make further predictions. 

\section{Formalism}

There are various approaches dealing with hadronic molecules as recently reviewed in Refs.~\citep{Guo:2017jvc,Chen:2016qju}. The LHG formalism seems working well to give a general consistent explanation for various observed hadronic molecular candidates with hidden charm~\cite{Dong:2021juy,Dong:2020hxe}. 
In Ref. \cite{Xiao:2019gjd}, using the CCUA with LHG formalism, we considered the coupled channels of the $J/\psi \Lambda$ channel, where there were nine channels of 
$\eta_c \Lambda$, $J/\psi \Lambda$, $\bar{D} \Xi_c$, $\bar{D}_s \Lambda_c$, $\bar{D} \Xi_c'$, $\bar{D}^* \Xi_c$, $\bar{D}^*_s \Lambda_c$, $\bar{D}^* \Xi_c'$, $\bar{D}^* \Xi_c^*$
in the $J^P=\frac{1}{2}^-,\ I=0$ sector, and six channels of 
$J/\psi \Lambda$, $\bar{D}^* \Xi_c$, $\bar{D}_s^* \Lambda_c$, $\bar{D}^* \Xi_c'$, $\bar{D} \Xi_c^*$, $\bar{D}^* \Xi_c^*$ 
in the $J^P=\frac{3}{2}^-,\ I=0$ sector. 
In addition, a single channel of $\bar{D}^* \Xi_c^*$ with $J^P=\frac{5}{2}^-$ was also found in $s$ wave within the HQSS, which was not taken into account in our work since it could not couple to the $J/\psi \Lambda$ channel in $s$ wave. 
Note that the interaction of the $\bar{D}^* \Xi_c^*$, which can be specified with spin $J=\frac{1}{2}, \, \frac{3}{2}, \, \frac{5}{2}$, was included in the coupled channels in Ref. \cite{Xiao:2019gjd} under the constraint of the HQSS, and not considered in Refs. \cite{Wu:2010jy,Wu:2010vk}. 

Within the CCUA, the scattering amplitudes ($T$) are evaluated by the coupled channel Bethe-Salpeter equation with the on-shell prescription,
\begin{equation}
T = [1 - V \, G]^{-1}\, V,
\label{eq:BS}
\end{equation}
where $G$ is constructed by the loop functions with meson-baryon intermediate states and $V$ are the potentials of the coupled channel interactions. 
Note that, $G$ is a diagonal matrix with elements of meson-baryon loop functions, where we take the ones with the dimensional regularization, see more details in Refs. \cite{Wu:2010jy,Wu:2010vk}. 
Thus, the free parameter is the only one of $a_\mu$, see the discussions later. 
Respecting the HQSS, the elements of the potential $V$ matrix are given in Tables~\ref{tab:vij11} and \ref{tab:vij31} for the $J=1/2, \, I=0$ and $J=3/2, \, I=0$ sectors, respectively, where we only show $V_{ij}$ for $j \geq i$ for simplicity due to the fact that $V_{ji}=V_{ij}$ in the CCUA. 
In Tables~\ref{tab:vij11} and \ref{tab:vij31}, the coefficients $\mu_{i}$, $\mu_{ij}$ ($i,j=1,2,3,4$) and $\lambda$ are the unknown low energy constants with the HQSS constraint, see more details in Ref. \cite{Xiao:2019gjd}. 

\begin{table}
     \renewcommand{\arraystretch}{1.7}
     \setlength{\tabcolsep}{0.2cm}
\centering
\caption{Potential matrix elements  $V_{ij}$ of Eq.~(\ref{eq:BS}) for the  $J=1/2,~I=0$ sector.}
\label{tab:vij11}
\begin{tabular}{ccccccccc}
\hline\hline
$\eta_c \Lambda$ & $J/\psi \Lambda$ &  $\bar D \Xi_c$ &  $\bar D_s \Lambda_c$ &  $\bar D \Xi'_c$ &  $\bar D^* \Xi_c$
  &  $\bar D^*_s  \Lambda_c$ &  $\bar D^* \Xi'_c$  &  $\bar D^* \Xi^*_c $   \\
\hline
$\mu_1$ & 0 & $-\frac{\mu_{12}}{2}$ & $-\frac{\mu_{13}}{2}$ & $\frac{\mu_{14}}{2}$ & $\frac{\sqrt{3}
   \mu_{12}}{2}$ & $\frac{\sqrt{3} \mu_{13}}{2}$ & $\frac{\mu_{14}}{2 \sqrt{3}}$ & $\sqrt{\frac{2}{3}} \mu_{14}$ \\    
  & $\mu_1$ & $\frac{\sqrt{3} \mu_{12}}{2} $& $\frac{\sqrt{3} \mu_{13}}{2}$ & $\frac{\mu_{14}}{2 \sqrt{3}}$
   & $\frac{\mu_{12}}{2}$ & $\frac{\mu_{13}}{2}$ & $\frac{5 \mu_{14}}{6}$ & $-\frac{\sqrt{2} \mu_{14}}{3}$ \\    
 &  & $\mu_2$ & $\mu_{23}$ & 0 & 0 & 0 & $\frac{\mu_{24}}{\sqrt{3}}$ & $-\sqrt{\frac{2}{3}} \mu_{24}$ \\  
  &  &  & $\mu_3$ & 0 & 0 & 0 & $\frac{\mu_{34}}{\sqrt{3}}$ & $-\sqrt{\frac{2}{3}} \mu_{34}$ \\  
 &  &  &  & $\frac{1}{3} (2 \lambda +\mu_4)$ &   $\frac{\mu_{24}}{\sqrt{3}}$ & $\frac{\mu_{34}}{\sqrt{3}}$ & $-\frac{2 (\lambda -\mu_4)}{3 \sqrt{3}}$ &
   $\frac{1}{3} \sqrt{\frac{2}{3}} (\mu_4-\lambda )$ \\    
 &  &  &  &  & $\mu_2$ & $\mu_{23}$ & $\frac{2 \mu_{24}}{3}$ & $\frac{\sqrt{2} \mu_{24}}{3}$ \\    
  &  &  &  &  &  &  $\mu_3$ & $\frac{2 \mu_{34}}{3}$ & $\frac{\sqrt{2} \mu_{34}}{3}$ \\  
 &  &  &  &  &  &  & $\frac{1}{9} (2 \lambda +7 \mu_4)$ & $\frac{1}{9} \sqrt{2} (\lambda -\mu_4)$ \\    
 &  &  &   &  &  &  &  & $\frac{1}{9} (\lambda +8 \mu_4)$ \\
\hline\hline
\end{tabular}
\end{table}

\begin{table}
     \renewcommand{\arraystretch}{1.7}
     \setlength{\tabcolsep}{0.2cm}
\centering
\caption{Potential matrix elements  $V_{ij}$ of Eq.~(\ref{eq:BS}) for the $J=3/2,~I=0$ sector.}
\label{tab:vij31}
\begin{tabular}{cccccc}
\hline\hline
 $J/\psi \Lambda$ & $ \bar D^* \Xi_c$ &  $\bar D_s^* \Lambda_c $
&  $\bar D^* \Xi'_c$  &  $\bar D \Xi^*_c$  & $\bar D^* \Xi_c^*$  \\
\hline
 $\mu_1$ & $\mu_{12}$ & $\mu_{13}$ & $-\frac{\mu_{14}}{3}$ & $\frac{\mu_{14}}{\sqrt{3}}$ &  $\frac{\sqrt{5} \mu_{14}}{3}$ \\    
  & $\mu_2$ & $\mu_{23}$ & $-\frac{\mu_{24}}{3}$ & $\frac{\mu_{24}}{\sqrt{3}}$ &  $\frac{\sqrt{5} \mu_{24}}{3}$ \\    
  &  & $\mu_3$ & $-\frac{\mu_{34}}{3}$ & $\frac{\mu_{34}}{\sqrt{3}}$ & $\frac{\sqrt{5} \mu_{34}}{3}$ \\   
 &  &  & $\frac{1}{9} (8 \lambda +\mu_4)$ &  $\frac{\lambda -\mu_4}{3 \sqrt{3}}$ & $\frac{\sqrt{5}}{9} (\lambda -\mu_4)$ \\   
 &  & &  & $\frac{1}{3} (2 \lambda +\mu_4)$ & $\frac{1}{3} \sqrt{\frac{5}{3}} (\mu_4-\lambda )$ \\   
 & &  &  &  & $\frac{1}{9} (4 \lambda +5 \mu_4)$ \\
\hline\hline
\end{tabular}
\end{table}

Using the LHG formalism, we obtain the values of these low energy constants \cite{Xiao:2019gjd},
\begin{eqnarray}
\mu_1 &=& \mu_3 =  \mu_{24} = \mu_{34} = 0\\ 
\mu_2 &=& \mu_{23}/\sqrt{2}= \mu_4= \lambda=-F, \quad
F=\frac{1}{4f^2} (p^0 + p^{\prime\, 0}) \\
\mu_{12} &=& -\mu_{13}/\sqrt{2}=\mu_{14}/\sqrt{3} =
-\sqrt{\frac{2}{3}}\ \frac{m_V^2}{m_{D^*}^2}\ F,
\end{eqnarray}
with $f_\pi = 93 \mev$ and $m_V = 800 \mev$, where $p^0$ and $p'\,^0$ are the energies of the incoming and outgoing mesons in a certain channel.
Note that, we have explicitly taken the reduction factor $m_V^2/m_{D^*}^2$ in the matrix elements that involve the transition processes with the exchange of $D^*$ meson. 
The null $\mu_{24}$ and $\mu_{34}$ values are due to the pion exchange neglect in our formalism, as done in Ref. \cite{Xiao:2013yca}, see more discussions in Ref. \cite{Xiao:2020frg}.

\section{Results and discussions}

As discussed above, the subtraction constant $a_\mu$ in the meson-baryon loop functions is a free parameter in our formalism, and thus, we cannot get a precise value for it theoretically.
It is even worse that the prediction will significantly rely on its value for a loose bound system, since its value will influence the lowest strength of attractive potential to form a bound state (we will discuss this issue in detail later).
In practice, the only way to get its accurate value is using some experimental data to fix it.
Therefore, using the newest experimental results of \cite{Aaij:2020gdg}, the value of $a_\mu$ can be determined as  $a_\mu (\mu=1\gev) = -1.94$ for the $P_{cs}$ case.
It is similar to what we have done in Ref.~\cite{Xiao:2019aya} for the $P_c$ case, where a value of $a_\mu (\mu=1\gev) = -2.09$ was obtained.
In the early prediction of Refs. \cite{Wu:2010jy,Wu:2010vk}, the central value of $a_\mu (\mu=1\gev) = -2.3$ was used to match the cutoff in the loop function as $0.8\gev$, which is close to the masses of exchange vectors, $\rho$ and $\omega$.  
Indeed, now the values of $a_\mu (\mu=1\gev) = -1.94$ and the one of $a_\mu (\mu=1\gev) = -2.09$ \cite{Xiao:2019aya} are really around the ``natural values" of $a_\mu=-2$ \cite{Oller:2000fj}.

\begin{figure}
\centering
\includegraphics[scale=0.6]{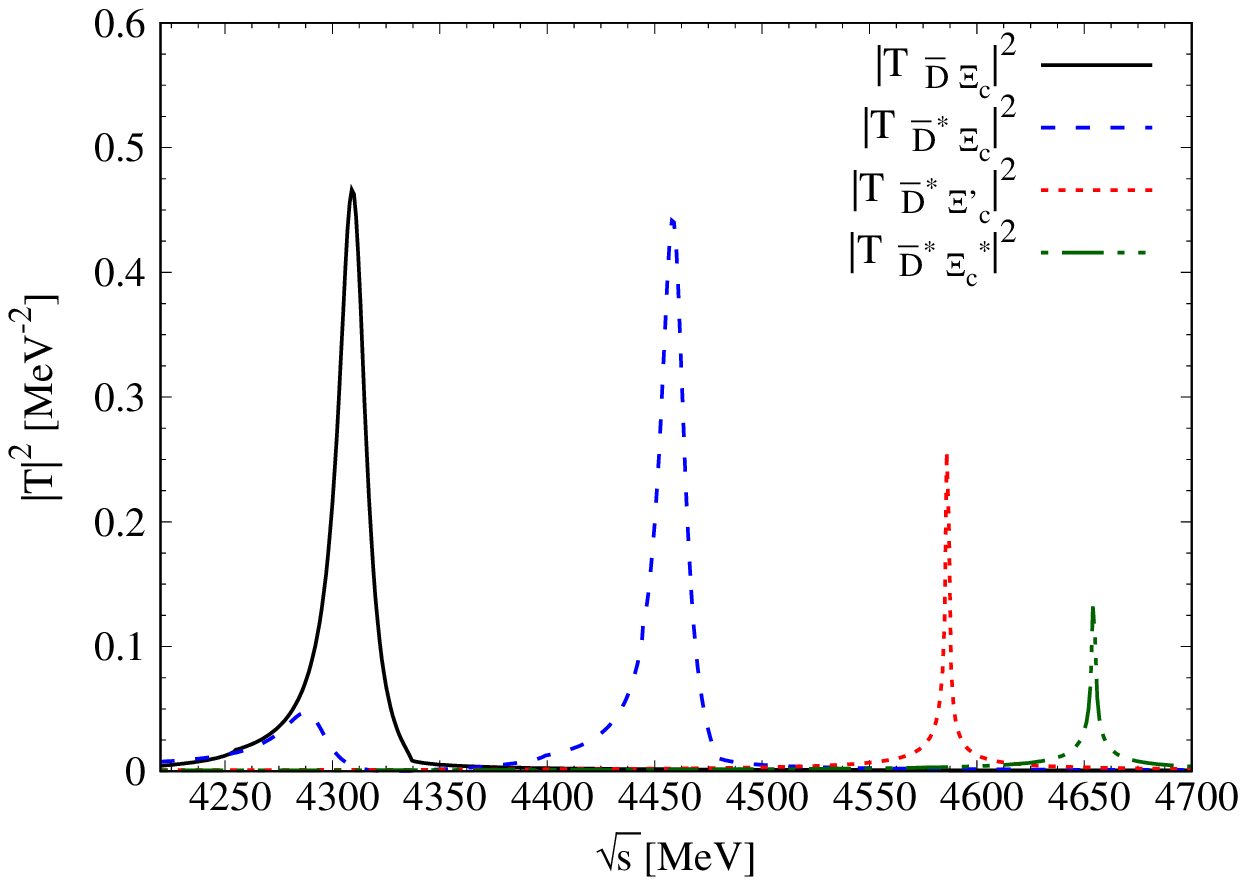}
\includegraphics[scale=0.6]{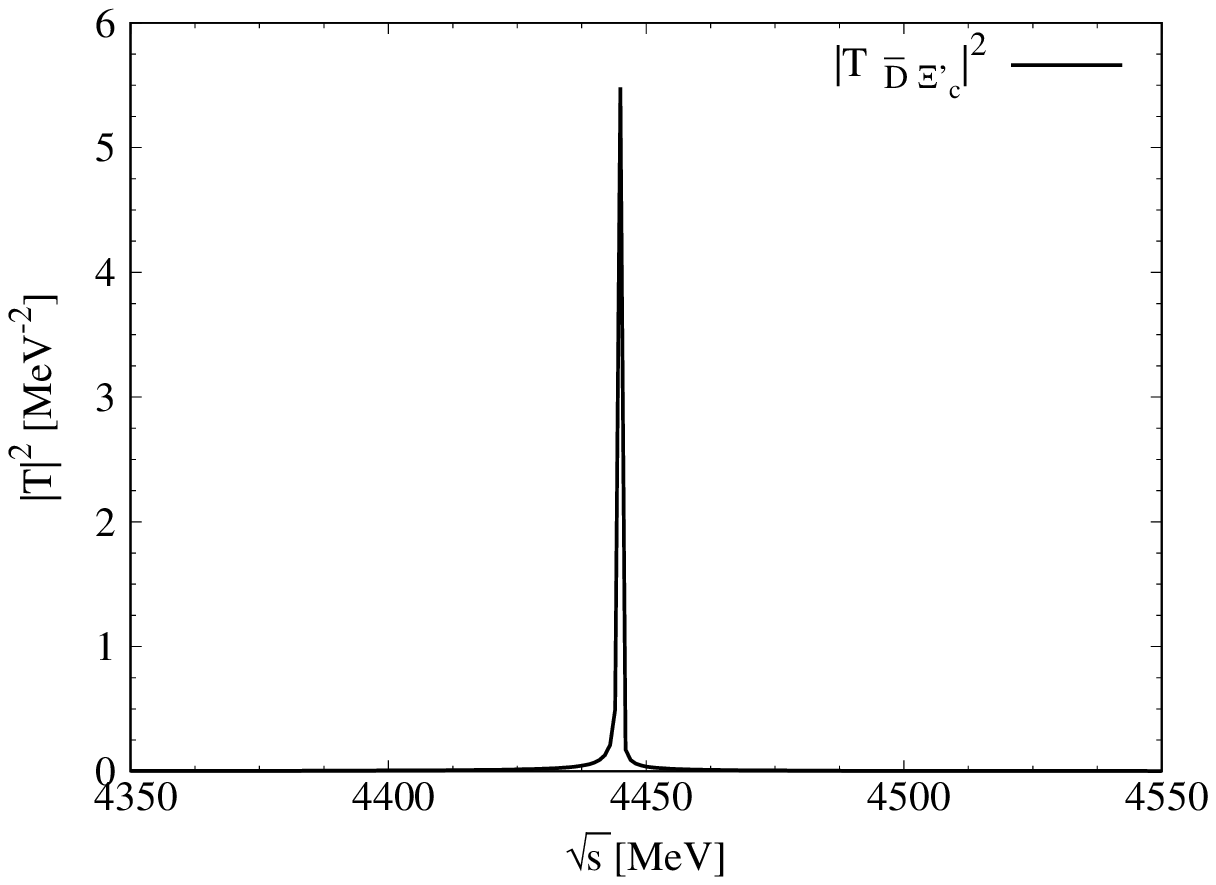}
\caption{Results of the modulus squared of the amplitudes for $J=1/2,~I=0$ sector.}
\label{fig:tsq12}
\end{figure}

\begin{figure}
\centering
\includegraphics[scale=0.6]{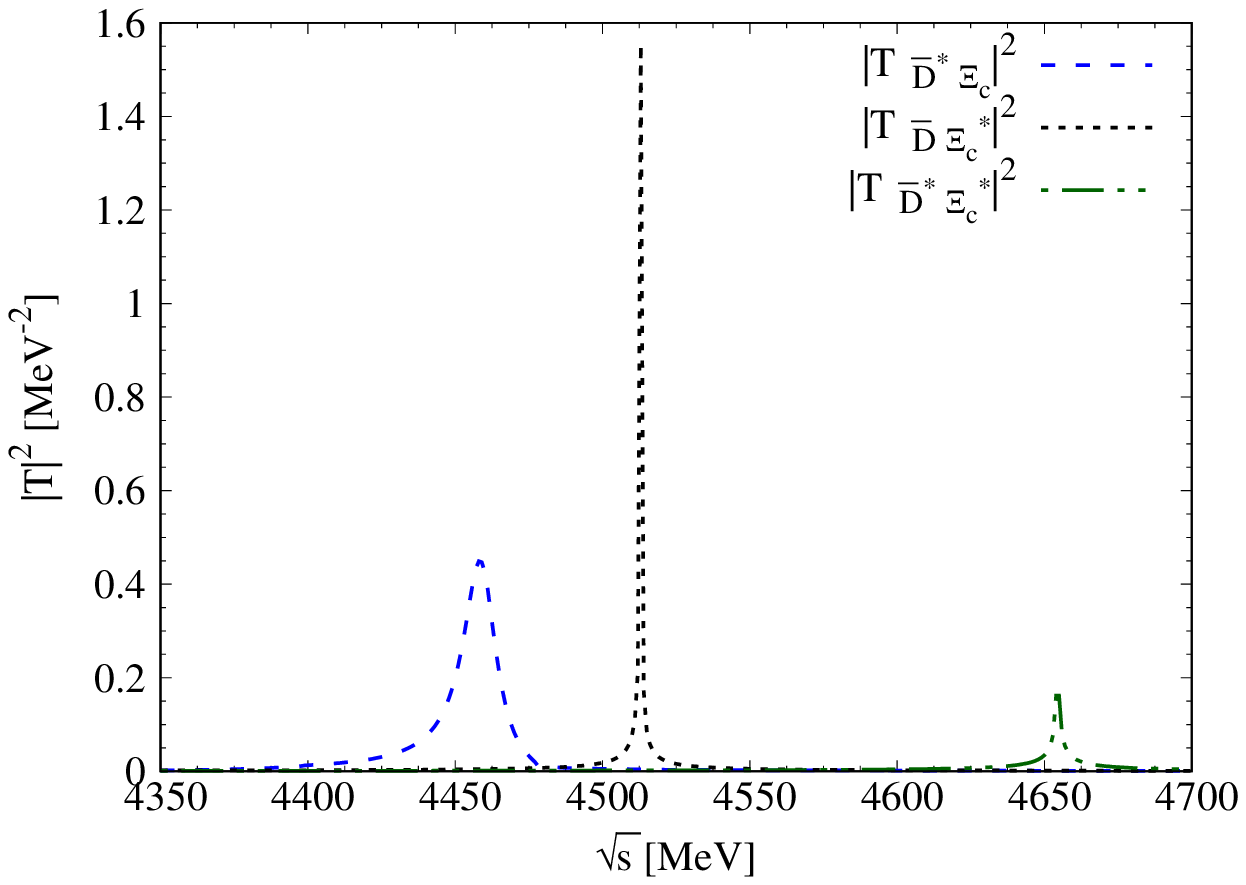}
\includegraphics[scale=0.6]{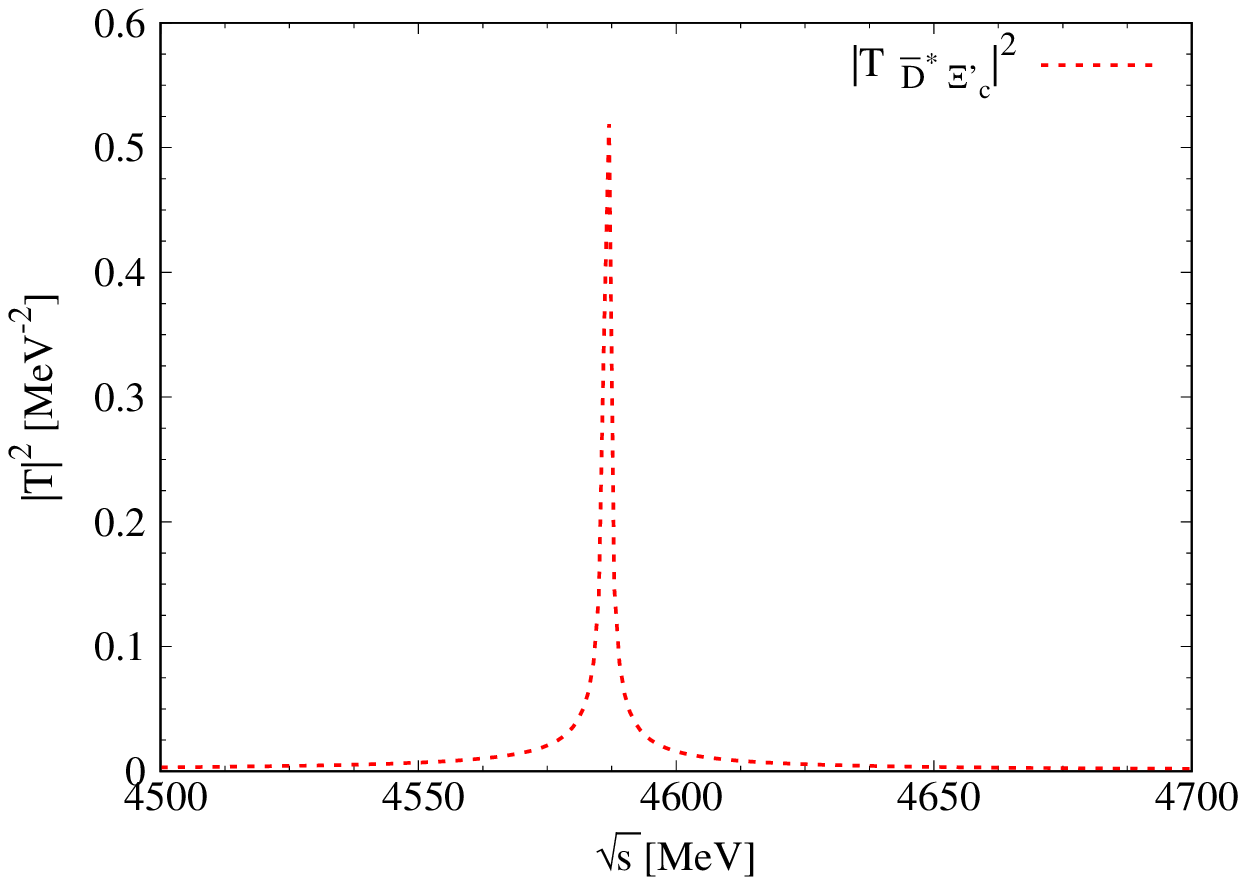}
\caption{Results of the modulus squared of the amplitudes for $J=3/2,~I=0$ sector.}
\label{fig:tsq32}
\end{figure}

Then, using the fitted $a_\mu (\mu=1\gev) = -1.94$,  we obtain the results of the modulus squared of the amplitudes in  Figs.~\ref{fig:tsq12} and ~\ref{fig:tsq32} for $J=1/2,~I=0$ and $J=3/2,~I=0$, respectively, where the peak structures are analogous to the ones obtained in Ref. \cite{Xiao:2019gjd} but more narrow and with higher energies. 
The corresponding poles and its coupling constants to all the channels are given in Tables~\ref{tab:cou12}
\footnote{There is an extra pole around $(4291.05+i 12.80) \mev$ in the $\bar D^* \Xi_c$ channel, see the left panel of Fig.~\ref{fig:tsq12}, which couples strongly to $\bar D^*_s  \Lambda_c$ channel (threshold $4398.66\mev$) and looks like unnormal. 
It is due to the fact that the $G$ functions for the channels of $\bar D^* \Xi_c$ and $\bar D^*_s  \Lambda_c$ (denoted as channel 6 and 7, respectively) become positive far below their thresholds, which also lead to the ``effective" potential of $V_{66} + V_{67}^2 G_{77}$ changing to a positive one (see the discussion later). Thus, a repulsive potential lead to a bound state unusually, see more discussions in Refs.~\cite{Xiao:2013yca,Wu:2010rv}.}  
and \ref{tab:cou32} for $J=1/2,~I=0$ and $J=3/2,~I=0$, respectively, where the thresholds of each channel are shown accordingly.
In Tables~\ref{tab:cou12} and \ref{tab:cou32}, with the couplings obtained, the partial decay widths and the branching ratios for each channel are evaluated.
One thing should be mentioned, that the poles given in Tables~\ref{tab:cou12} and \ref{tab:cou32} are the ones located in the general ``second Riemann sheet", which means that all the channels below the certain bound channel are extrapolated to the second Riemann sheet (these channels are always called the open channels), whereas, the other coupled channels including the bound channel are in the first Riemann sheet. 
As found from our results of Tables~\ref{tab:cou12} and \ref{tab:cou32}, the bound state of $\bar D^* \Xi_c$ is tuned as $4459 \mev$ to make it consistent with the mass of observed $P_{cs}$ state, which has increased about 30 MeV with respect to the one obtained in Ref.~\cite{Xiao:2019gjd}. 
Then the pole of the $\bar D^* \Xi_c$ channel is $(4459.07+i6.89)\mev$, where the width is quite consistent with the experimental results~\cite{Aaij:2020gdg}, and it just has 3 MeV difference.
Note that, owing to the pion exchange neglect in our formalism as discussed above, this pole of the $\bar D^* \Xi_c$ channel is degenerate with spins $J=1/2$ and $J=3/2$, as shown in Tables~\ref{tab:cou12} and \ref{tab:cou32}.
Therefore, one can conclude from our formalism that the $P_{cs} (4459)$ state can be a bound state of $\bar D^* \Xi_c$ with spin uncertainty of $J=1/2$ or $J=3/2$. 
Besides, in Table~\ref{tab:cou12} for the $J=1/2,~I=0$ sector, we have another very stable pole, $(4310.53+i8.23)\mev$, which is bound by the $\bar D \Xi_c$ channel and has 56 MeV binding energy.
On the other hand, other three peak structures are all very close to the corresponding thresholds.
The peak around 4445 MeV is contributed from the pole at $(4445.12+i0.19)\mev$ bound by the $\bar D \Xi'_c$ channel, of which the binding energy is just 0.23 MeV. 
Thus, in our model this bound state becomes unstable if the parameter $a_\mu$ changes a little to move it to the threshold, which will cause the threshold effect and then lead to no pole in the general second Riemann sheet, as shown in Tables~\ref{tab:cou12} and \ref{tab:cou32} for some channels.
Furthermore, the other two peak structures are actually due to the poles from the other Riemann sheets.
Thus, these poles can not be recognized as normal bound states of the certain channels.
It is proper to say that these two peak structures are the threshold effects because of very weak attractive interaction potentials. 
Indeed, the three poles for the channels of $\bar D \Xi'_c$, $\bar D^* \Xi'_c$ and $\bar D^* \Xi^*_c$ were just loosely bound as found in the results of Ref.~\cite{Xiao:2019gjd}, where these poles were only a few MeV below the corresponding thresholds and had narrow widths, a few MeV.
Similarly, in Table~\ref{tab:cou32} for $J=3/2,~I=0$ sector, except for the one of $\bar D^* \Xi_c$ stable, the other three poles of the channels $\bar D^* \Xi'_c$, $\bar D \Xi^*_c$ and $\bar D^* \Xi_c^*$ are not stable for the same reason. 
Three of them were also loosely bound as given in the results of Ref.~\cite{Xiao:2019gjd} with a few MeV for the binding energies and the widths. 

\begin{figure}
\centering
\includegraphics[scale=0.6]{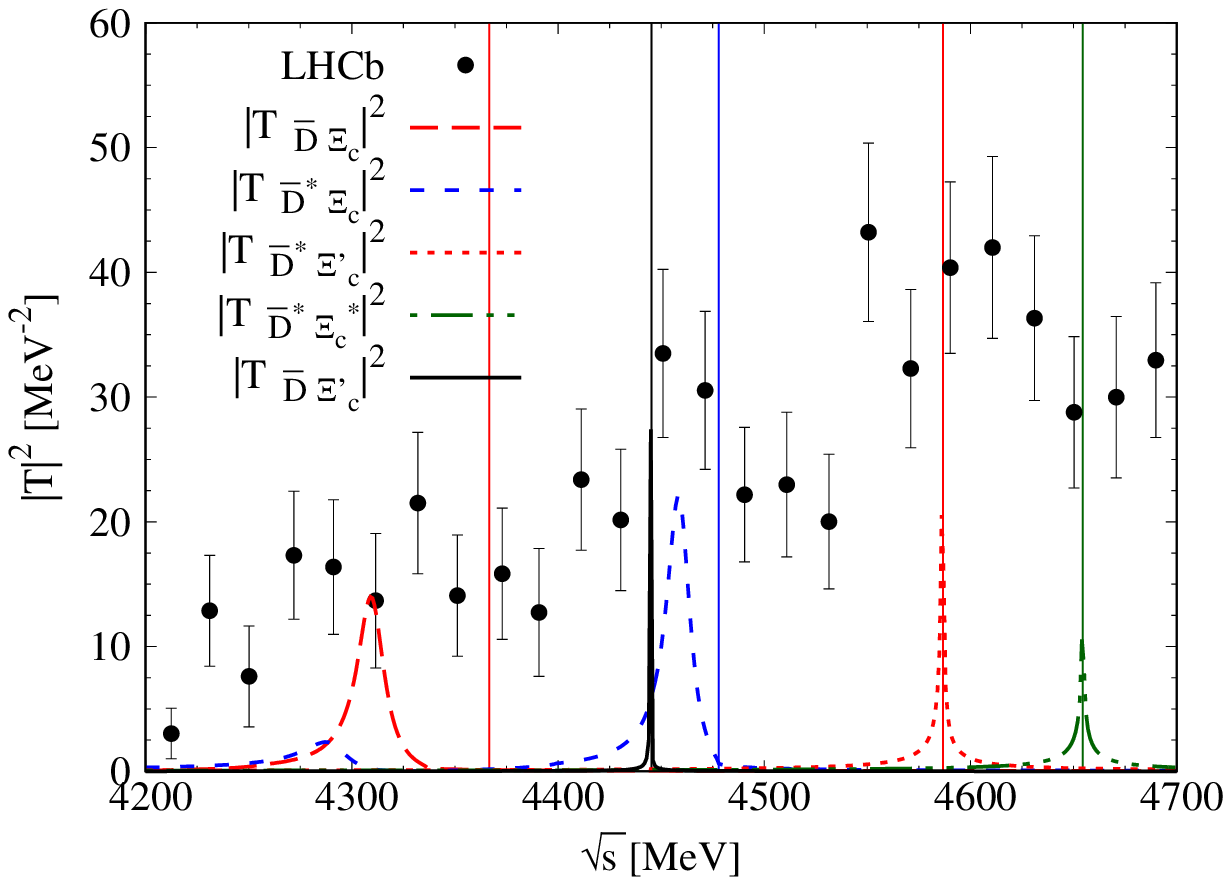}
\includegraphics[scale=0.6]{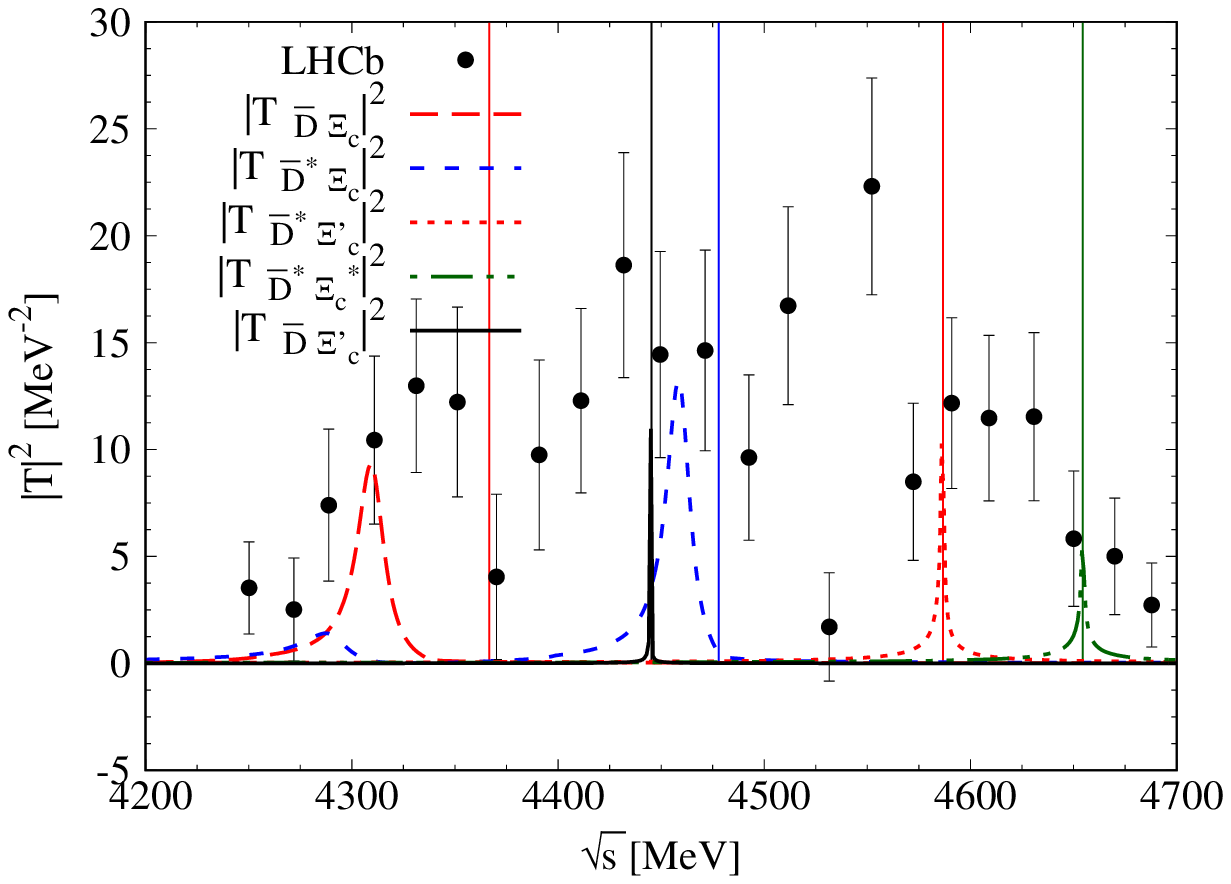}
\caption{Results of the modulus squared of the amplitudes for $J=1/2,~I=0$ sector compared with the experimental data, Left: with full data; Right: with $1.9\gev < m_{\Lambda K^-} < 2.2\gev$. The vertical lines correspond to the thresholds of the certain channels.}
\label{fig:tsqcom}
\end{figure}

As shown in Tables~\ref{tab:cou12} and \ref{tab:cou32}, the degenerate molecular states of $\bar D^* \Xi_c$ with $J=1/2$ and $J=3/2$ correspond to nearly the same pole $(4459.07+i6.89)\mev$. 
Their main decay channel is $\bar D^*_s  \Lambda_c$ with the branch ratio larger than $80\%$, which can be understood as this channel coming from the strong interaction by exchanging light vector meson in our model, and its large decay width was also found in Ref.~\cite{Chen:2021tip}.
But, the branch ratios of the $J/\psi \Lambda$ channel in the two cases are a little different because of the Clebsch-Gordan coefficients in the HQSS model.
Besides, the $\eta_c \Lambda$ channel is the only decay channel of the bound state with $J^P=1/2^-$, while it requires the $d$ wave interaction for the $J^P=3/2^-$ state, which is neglected in our model.
Therefore, if we assume the $\bar D^* \Xi_c$ bound state to be the $P_{cs}(4459)$ state, it may have two-poles structure with different spins, like the one of $P_c(4450)$ before, and it can be helpful to reveal their nature with more experimental information in the future for the partial decay widths and the branching ratios of the channels $J/\psi \Lambda$ and $\eta_c \Lambda$.
In Fig.~\ref{fig:tsqcom}, we show our results of Fig.~\ref{fig:tsq12} for $J=1/2,~I=0$ compared with the experimental data, where the full data are compared on the left and the parts with the cut of $1.9\gev < m_{\Lambda K^-} < 2.2\gev$ on the right. 
As one can see from Fig.~\ref{fig:tsqcom}, at the present Run II data, there is not enough statistics to draw a strong conclusion for the existence of more bound states in the $J/\psi \Lambda$ invariant mass distributions.

\begin{table}
     \renewcommand{\arraystretch}{1.3}
     \setlength{\tabcolsep}{0.2cm}
\centering
\caption{Coupling constants to all channels for certain poles in $J=1/2,~I=0$ sector.} \label{tab:cou12}
\begin{tabular}{ccccc ccccc}
\hline\hline
Chan.  & $\eta_c \Lambda$  & $J/\psi \Lambda$  &  $\bar D \Xi_c$  &  $\bar D_s \Lambda_c$  &  $\bar D \Xi'_c$  
  &  $\bar D^* \Xi_c$  & $\bar D^*_s  \Lambda_c$  & $\bar D^* \Xi'_c$   &  $\bar D^* \Xi^*_c$    \\
Thres.  & 4099.58 & 4212.58 & 4366.61 & 4254.80 & 4445.34 & 4477.92 & 4398.66 & 4586.66 & 4654.48  \\
\hline
\multicolumn{2}{c}{$4310.53+i8.23$}  \\
\hline
$|g_i|$ & $0.15$ & $0.27$ & $\mathbf{2.33}$ & $0.69$ & $0.00$ & $0.04$ & $0.09$ & $0.01$ & $0.02$  \\
$\Gamma_i$ & $0.57$ & $1.18$ & $-$ & $13.86$ & $-$ & $-$ & $-$ & $-$ & $-$  \\
Br. & 3.47\% & 7.16\% & $-$ & 84.21\% & $-$ & $-$ & $-$ & $-$ & $-$  \\
\hline
\multicolumn{2}{c}{$4445.12+i0.19$}  \\
\hline
$|g_i|$ & $0.10$ & $0.06$ & $0.00$ & $0.00$ & $\mathbf{0.72}$ & $0.08$ & $0.04$ & $0.01$ & $0.01$  \\
$\Gamma_i$ & $0.29$ & $0.08$ & $0.00$ & $0.00$ & $-$ & $-$ & $0.04$ & $-$ & $-$  \\
Br. & 74.74\% & 21.22\% & 0.01\% & 0.01\% & $-$ & $-$ & 10.62\% & $-$ & $-$  \\
\hline
\multicolumn{2}{c}{$4459.07+i6.89$}  \\
\hline
$|g_i|$ & $0.22$ & $0.13$ & $0.00$ & $0.00$ & $0.07$ & $\mathbf{2.16}$ & $0.61$ & $0.03$ & $0.02$  \\
$\Gamma_i$ & $1.59$ & $0.46$ & $0.00$ & $0.00$ & $0.01$ & $-$ & $11.14$ & $-$ & $-$  \\
Br. & 11.57\% & 3.31\% & 0.00\% & 0.00\% & 0.70\% & $-$ & 80.86\% & $-$ & $-$  \\
\hline
\multicolumn{2}{c}{$4586.66?$}  \\
\hline
$|g_i|$ & $-$  &   $-$  &  $-$  &  $-$  &  $-$  & $-$  & $-$  & $-$  & $-$   \\
\hline
\multicolumn{2}{c}{$4654.48?$}  \\
\hline
$|g_i|$ & $-$  &  $-$  &  $-$ &  $-$  &  $-$  & $-$  & $-$  & $-$  & $-$   \\
\hline
\end{tabular}
\end{table}

\begin{table}
     \renewcommand{\arraystretch}{1.3}
    \setlength{\tabcolsep}{0.3cm}
\centering
\caption{Coupling constants to all channels for certain poles in $J=3/2,~I=0$ sector..} \label{tab:cou32}
\begin{tabular}{cccc ccc}
\hline\hline
Chan. & $J/\psi \Lambda$ &  $\bar D^* \Xi_c$ &  $\bar D_s^* \Lambda_c$ 
&  $\bar D^* \Xi'_c$  &  $\bar D \Xi^*_c$  & $\bar D^* \Xi_c^*$  \\
Thres. & 4212.58 & 4477.92 & 4398.66 & 4586.66 & 4513.17 & 4654.48 \\
\hline
\multicolumn{2}{c}{$4459.02+i6.83$}  \\
\hline
$|g_i|$ & $0.28$ & $\mathbf{2.16}$ & $0.61$ & $0.02$ & $0.04$ & $0.03$  \\
$\Gamma_i$ & $2.00$ & $-$ & $11.15$ & $-$ & $-$ & $-$  \\
Br.  & 14.68\% & $-$ & 81.64\% & $-$ & $-$ & $-$  \\
\hline
\multicolumn{2}{c}{$4586.66?$}  \\
\hline
$|g_i|$ & $-$  & $-$  & $-$  & $-$  & $-$  & $-$    \\
\hline
\multicolumn{2}{c}{$4513.17?$}  \\
\hline
$|g_i|$ & $-$  & $-$  & $-$  & $-$  & $-$  & $-$    \\
\hline
\multicolumn{2}{c}{$4654.48?$}  \\
\hline
$|g_i|$ & $-$  & $-$  & $-$  & $-$  & $-$  & $-$    \\
\hline
\end{tabular}
\end{table}

For the results of the partial decay widths and the branching ratios shown in Table~\ref{tab:cou12}, it seems strange that the bound state of $\bar D \Xi_c$ with its pole at $(4310.53+i8.23)\mev$ decays more strongly to the $J/\psi \Lambda$ channel than to the $\eta_c \Lambda$, whereas, the molecular state of $\bar D^* \Xi_c$ with the pole $(4459.07+i6.89)\mev$ is just the opposite, which is different from the cases of the $\bar{D}^{(*)} \Sigma_c$ bound states studied in Ref.~\cite{Xiao:2020frg}. 
This is because of the constraint of the HQSS, see the transition elements of $\frac{\mu_{12}}{2}$ and $\frac{\sqrt{3} \mu_{12}}{2}$ for the related channels in Table~\ref{tab:vij11}. 
But, in Ref.~\cite{Shen:2019evi} the bound state of $\bar{D}^* \Xi_c$  predicted in Refs.~\cite{Wu:2010jy,Wu:2010vk} was found to decay stronger to the $J/\psi \Lambda$ channel than the $\eta_c \Lambda$. 
Indeed, the earlier work of~\cite{Wu:2010jy,Wu:2010vk,Shen:2019evi} did not consider the HQSS as well as the transition between Pseudoscalar-Baryon (PB) and Vector-Baryon(VB) channels.
However, after taking into account the HQSS, the interactions between PB and VB are almost the same except for the Clebsch-Gordan coefficients.
On the other hand, the bound states of $\bar D \Xi_c$ and $\bar D^* \Xi_c$ decay mostly to the channels of $\bar D_s \Lambda_c$ and $\bar D^*_s  \Lambda_c$, respectively, which have quite large partial decay widths and the branching ratios (see the results in Tables~\ref{tab:cou12} and \ref{tab:cou32}), whereas, the cases of the $\bar{D}^{(*)} \Sigma_c$ bound states~\cite{Xiao:2020frg} cannot decay to the channels $\bar{D}^{(*)} \Lambda_c$ with the vector meson exchange. 
The reason is that the $\Lambda_c$ particle is in the same spin 1/2 antitriplet (the $\bar{3}$ multi-states) with the $\Xi_c$ particle in the SU(3) flavour symmetry, but not with the $\Sigma_c$ particle (belonging to the $6$ multi-states). 

\begin{figure}
\centering
\includegraphics[scale=0.85]{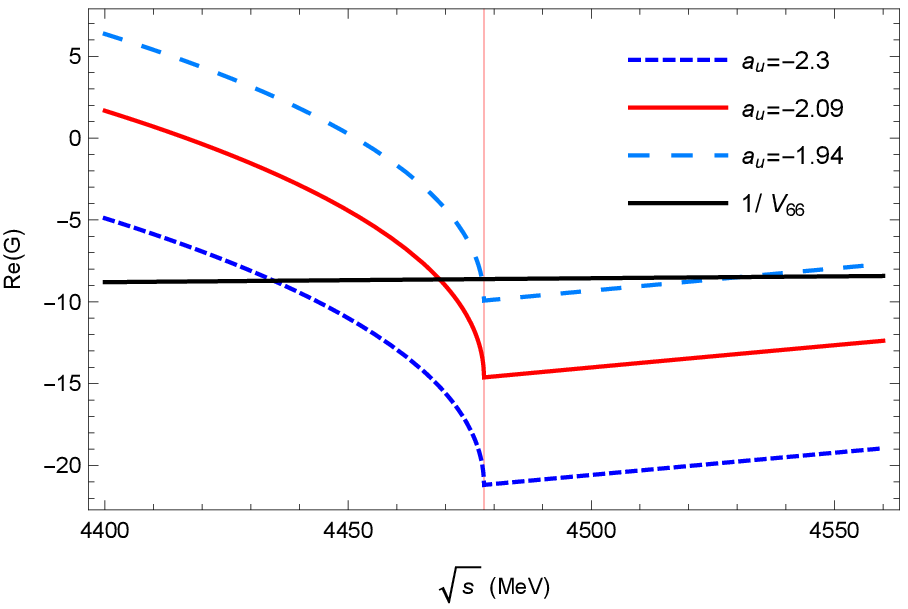}
\includegraphics[scale=0.85]{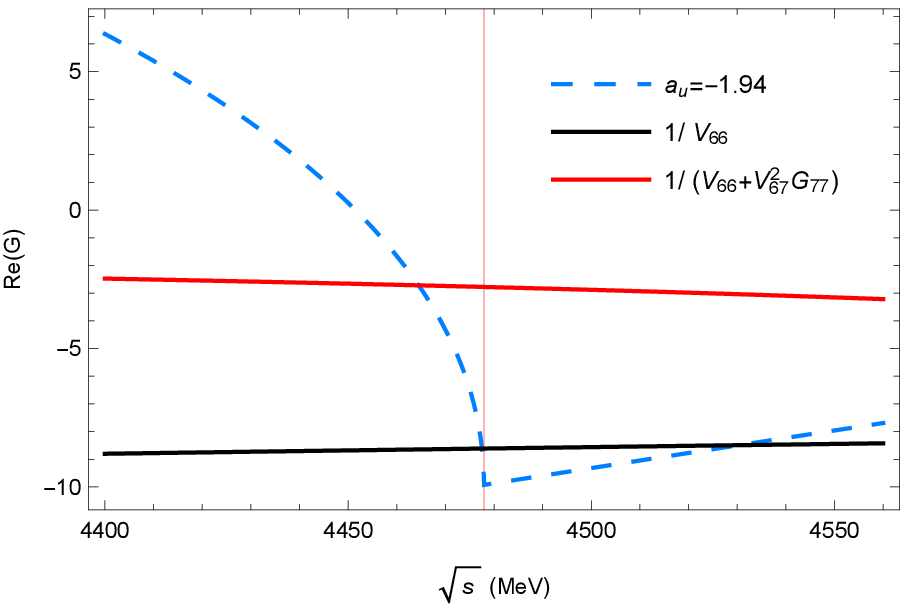}
\caption{Results of the potential $\frac{1}{V}$ versus the loop functions for the $\bar D^* \Xi_c$ channel, where the vertical line locates at the threshold.}
\label{fig:VGcom}
\end{figure}

As discussed above, when we take the new value of $a_\mu (\mu=1\gev) = -1.94$ for the only free parameter in the loop functions, the loose bound systems of $\bar D \Xi'_c$, $\bar D^* \Xi'_c$, $\bar D^* \Xi^*_c$ in $J=1/2,~I=0$ sector, and $\bar D^* \Xi'_c$, $\bar D \Xi^*_c$, $\bar D^* \Xi_c^*$ in $J=3/2,~I=0$ sector have become unstable with the poles moving to the thresholds, except for three strongly bound ones of $\bar D \Xi_c$ and $\bar D^* \Xi_c$. 
Note that, all of them were stable bound states in the results of Ref.~\cite{Xiao:2019gjd} with $a_\mu (\mu=1\gev) = -2.09$, and even more bound with $a_\mu (\mu=1\gev) = -2.3$ used in Refs.~\cite{Wu:2010jy,Wu:2010vk}, taking the one of $\bar D^* \Xi_c$ for example, the pole at 4370 MeV, which had been below the threshold of the $\bar D^*_s  \Lambda_c$ channel.
Indeed, the reduced $a_\mu$ leads to all of the poles less bound, and thus, some of the loosely bound states moved to the thresholds, see the left part of Fig.~\ref{fig:VGcom}, where we plot the real parts of the loop functions $G$ with different $a_\mu$ and the inverse potential $\frac{1}{V}$, taking the $\bar D^* \Xi_c$ channel for example (denoted as the channel 6 in the figure), see more discussions about the pole moving by the free parameter of the loop functions in Refs. \cite{Sekihara:2012xp,Ahmed:2020kmp}. 
Furthermore, there is a strong coupling between the $\bar D^*_s  \Xi_c$ (the channel 6 in the figure) and $\bar D^*_s  \Lambda_c$ (the channel 7 in the figure) channels, and then such off-diagonal transition potential will strengthen the attractive interaction from pure diagonal potential $V_{66}$ to the ``effective" potential of $V_{66} + V^2_{67} G_{77}$ owing to $V_{77} = \mu_3 = 0$ (see Table~\ref{tab:vij11}) .
In the right panel of Fig.~\ref{fig:VGcom}, we compare the real parts of the loop functions $G$ with the one of $1 / (V_{66} + V^2_{67} G_{77})$ (taking $a_\mu = -1.94$), where the cross point is just the pole position.  
Thus, one can expect that the $\bar D^* \Xi_c$ system will be more bound when $a_\mu (\mu=1\gev) = -2.3$ are taken as in Refs. \cite{Wu:2010jy,Wu:2010vk}, which is analogous to the case of the $\bar D \Xi_c$ system with the contribution from the $\bar D_s  \Lambda_c$ channel .

\section{Conclusion}

In summary, we revisit the interactions of the $J/\psi \Lambda$ channel and its coupled channels in $s$ wave, using the CCUA in combination with HQSS and LHG symmetry. 
With the new observation of the $P_{cs} (4459)$ state by LHCb to fix the only free parameter, we obtain a pole of $(4459.07+i6.89)\mev$ located below the threshold of the $\bar D^* \Xi_c$ channel, which represents effectively the two nearly degenerate states with spin parity $J^P = \frac{1}{2}^-$ and $J^P = \frac{3}{2}^-$ in our approach, which keeps in the potential only the constant leading term by dropping the momentum dependent terms. 
To remove the degeneracy, we will make further investigations in the future by taking into account the momentum dependent terms, including the pion exchange mechanism.
Thus, by assuming the LHCb observed $P_{cs} (4459)$ peak to be the degenerated $\bar D^* \Xi_c$ molecular states from our formalism, it would split to two peaks with higher statistics in the future, like the previous $P_c (4450)$ peak observed by LHCb.
From our results, there is also a $\bar D \Xi_c$ bound state with $J^P = \frac{1}{2}^-$ and a pole at $(4310.53+i8.23)\mev$, which is consistent with the predictions of Ref.~\cite{Chen:2020uif} within the uncertainties. 
Owing to the uncertainties of the experimental data and hence their corresponding constraint to the free parameter, the possible loosely bound states of $\bar D \Xi'_c$, $\bar D^* \Xi'_c$, $\bar D^* \Xi^*_c$ with $J=1/2,~I=0$ and $\bar D^* \Xi'_c$, $\bar D \Xi^*_c$, $\bar D^* \Xi_c^*$ with $J=3/2,~I=0$, which were predicted previously, are in fact suffering a large model dependence, since there are no coupled channels to strengthen the attractive interaction. 
Thus, unfortunately even their existence is now put into question in our present work. Although their LHG potentials are attractive, whether they are bound states or virtual states would sensitively depend on the model parameter and neglected momentum dependent terms.  These results are consistent with recent general analysis in Ref.~\citep{Dong:2021juy}. 
To look for these molecular states, especially for the $\bar D \Xi_c$ bound state and two $\bar D^* \Xi_c$ states (corresponding to the observed $P_{cs} (4459)$ peak), the decay channels of $\bar D_s  \Lambda_c$ and $\bar D^*_s  \Lambda_c$, respectively, are strongly suggested due to their large decay branching ratios. 
Also searching for them both in the $J/\psi \Lambda$ and $\eta_c \Lambda$ channels can be helpful to distinguish their different nature. 
We hope that future experiments, the Run-3 in LHCb for example, can make further test on our predictions and suggestions to reveal the properties of the $P_{cs}$ states.

\section*{Acknowledgments}

We thank useful discussions and valuable comments from Bo Fang about the experimental information, 
and acknowledge Eulogio Oset for useful comments and careful reading the paper. 
This work is partly supported by the Fundamental Research Funds for the Central Universities (J.J.Wu), 
and NSFC under Grant No. 12070131001 (CRC110 cofunded by DFG and NSFC), Grant No. 11835015, No. 12047503, and by the Chinese Academy of Sciences (CAS) under Grants No. XDB34030000 (B. S. Zou).

\end{document}